**Rule based Complex Event Processing for an Air Quality Monitoring System in Smart City**

Shashi Shekhar Kumar[a*], Ritesh Chandra[b], Sonali Agarwal[c]

Indian Institute of Information Technology Allahabad, India

rsi2020502@iiita.ac.in, rsi2022001@iiita.ac.in, sonali@iiita.ac.in

**Abstract**

In recent years, smart city-based development has gained momentum due to its versatile nature in architecture and planning for the systematic habitation of human beings. According to World Health Organization (WHO) report, air pollution causes serious respiratory diseases. Hence, it becomes necessary to real-time monitoring of air quality to minimize effect by taking time-bound decisions by the stakeholders. The air pollution comprises various compositions such as $NH_3$, $O_3$, $SO_2$, $NO_2$, etc., and their concentrations vary from location to location. The research work proposes an integrated framework for monitoring air quality using rule-based Complex Event Processing (CEP) and SPARQL queries. CEP works with the data stream based on predefined rules to detect the complex pattern, which helps in decision support for stakeholders. Initially, the dataset was collected from the Central Pollution Control Board (CPCB) of India and this data was then preprocessed and passed through Apache Kafka. Then a knowledge graph developed based on the air quality paradigm. Consequently, convert preprocessed data into Resource Description Framework (RDF) data, and integrate with Knowledge graph which is ingested to CEP engine using Apache Jena for enhancing the decision support . Simultaneously, rules are extracted using a decision tree, and some ground truth parameters of CPCB are added and ingested to the CEP engine to determine the complex patterns. Consequently, the SPARQL query is used on real-time RDF dataset for fetching the condition of air quality as good, poor, severe, hazardous etc based on complex events detection.  For validating the proposed approach various chunks of RDF are used for the deployment of events to the CEP engine, and its performance is examined over time while performing simple and complex queries.

1. **Introduction**

In recent years, smart cities have surfaced as a core area of research for developing future-ready, sustainable cities leveraging planned architecture to provide a healthy life for human beings. The "smart city" is termed as an urbanized area integrated with emerging technologies to collect information from local citizens and implement various measures for uplifting their standards of living[1]. These operations include Information and Communication Technology (ICT), the Internet of Things (IoT), Machine Learning (ML), Big Data and CEP for effective decision-making and promoting the technologies for more efficient interaction among the stakeholders[1].

According to World Bank data till 2022, the urban population has increased by up to 57% and in the Indian context, the value is up to 35.87% [2]. The latest report of the United Nations predicts that by the year 2050, it is expected that 68% [3] of world population will be residing in urban cities. The growth and development of smart cities comes up with major challenges such as health care, air quality, sustainability and environmental issues.

As per the records of the WHO, due to air pollution, around 6.7 million people die prematurely annually from health complications, which are widespread around the globe. The 3.2 million deaths are caused by urban air pollution, which includes 2.37 lakh children, as per data from 2020 [4]. Apart from global data, the Lancet report says that in 2019, air pollution was responsible for 1.67 millions (95% confidence interval: 1.42 - 0.92) fatalities, or 17.8% (15.8 - 19.5) of all deaths in the nation. The majority of these fatalities (98 million [0.77-1.19]) were caused by residential air pollution and 61 million [0.39-0.86] by ambient Particulate Matter (PM) pollution [5].

Particulate matter is an important substance among all air pollutants and is measured in terms of its size (2.5µm) or $PM_{2.5}$. The monitoring of pollutants ($SO_2$, $NO_X$, $O_3$, $NO_2$, $NH_3$, CO, NO, $PM_{2.5}$, $PM_{10}$) is done twice a week for a duration of 24 hours (4 hours for gaseous pollutants and 8 hours for particulate matter), resulting in a total of 104 observations in a year.[2] Furthermore, for each pollutant substance, there are some specific permissible values; if these values exceed them, we define them with PM values and classify them according to the standard defined by the Govt and perform the required action to mitigate them to maintain air quality [6].

When it pertains to smart cities in the current domain, numerous pieces of work have been done so far, and it is still very much open for further research using various emerging technologies that can improve prediction and monitoring efficiently and reliably. In recent times, CEP has emerged as an important technology to determine the useful patterns from streaming data and extract useful information that makes the decision process easier. The CEP-based smart air quality monitoring system is highly beneficial as it utilizes rule-based filtering of event streams, correlates them for analysis of a huge amount of data, and establishes windowing for the extraction of information with in a specified time frame [7].

---

[1] https://smartcities.gov.in/
[2] https://cpcb.nic.in/about-namp/

Based on the motivation of previous studies and current research work, we propose this work in two different phases. In the first phase, a dataset is collected from smart city air quality data, which includes attributes such as $PM_{10}$, $PM_{2.5}$, Ammonia ($NH_3$), Sulphur Dioxide ($SO_2$), Carbon Monoxide (CO), inhalable particles ($PM_{10}$), Nitrogen Monoxide (NO), Nitrogen Oxide (NOx), Nitrogen Dioxide ($NO_2$), and Ozone ($O_3$).

The data is preprocessed to remove missing and ambiguous values from the stored database. Later, the dataset is streamed through Apache Jena and passed through Apache Kafka for event processing. Subsequently, rules are generated using decision trees, and additionally, we have added standard parameters for rules authorized by the CPCB, Govt. of India.

Analogously, in the second phase, preprocessed data is converted into RDF data using GraphDB and combined with a knowledge graph for extracting additional information related to air quality. Then ingest RDF data in the CEP engine, which continuously monitors and filters the data streams for complex events using predefined rules. Subsequently, we perform a query based on standard parameters defined by CPCB to know the current scenario of Air Quality Data using SPARQL through GraphDB and Apache Jena. The queries are tested on various chunks of the RDF dataset, and the execution time is observed for efficient query processing. We also examine the number of events deployed through the CEP engine to perform an evaluation of how efficiently our proposed work is able to handle queries.

**The main contribution of this research work followed as :**
1. To design an integrated framework for performing event deployment from data streams through CEP Engine.
2. To extract rules through decision trees and standard parameters certified by CPCB.
3. Efficient query can be processed on simple and complex queries based on different RDF datasets.
4. Getting real time insights of Air Quality Datastream with purely environmental perspective in a smart city.
5. Performance of events processing through Siddhi CEP Engine based on execution time of different chunks of data by using the proposed approach.

6. A CEP based decision support system (DSS) for Air Quality Monitoring.

The rest of this paper is organized as follows: Section **II** gives a glimpse about the related work presented in this domain. Section **III** presents the preliminary requirements and working of the proposed work. Section **IV** presents experimental results details for this research work. Section **V** discusses conclusion and future work.

## 2 Literature Review

In this section, we have explored the recent work done in this domain and how research has gone with smart city air quality monitoring. There are numerous studies available in the context of air quality prediction and monitoring for efficient management. Various technologies and methodologies have been implemented, but some areas still need exploration. Through this literature survey, we present a detailed analysis of how they lack this current work.

### 2.1 Air Quality Monitoring for Smart City

Chatti et al. [8] proposed a smart long-range-based node for timely collection of air quality data and updating its cloud for efficient monitoring. It used a long-range, wide-range network (LoRaWAN-IoT-AQMS) for air quality monitoring systems (AQMS) with the deployment of models in an open environment. They integrated the proposed work with a real-time web dashboard and compared the work with experimental results, which shows that air quality indicators are well monitored.

Martín-Baos et al. proposed a low-cost IoT-based system for monitoring traffic-based air quality index (AQI). The traffic flow was analyzed using compressed video and an AQI index supported by machine learning and regression techniques using different feature data. The experimental results were evaluated using various performance metrics precision and recall that improved the performance of AQI [9].

Kortoçi et. al. designed a low-cost portable sensor to demonstrate citizen-based pollution measurements based on indoor and outdoor data. Through their experimentation, they validated the consistency and accuracy of the proposed work by identifying probable causes of air pollution dependent on geographical areas. Their model also gives personalized insights about air pollutants [10].

Cui et. al. [11] proposed a study based on how efficiently smart cities lower $PM_{2.5}$ in the environment in smart cities in China. They found that $PM_{2.5}$ has been reduced by 5.34 % in specific ranges of city premises. The reducing effect of smart cities is primarily based on technology upgrades and promoting various aspects of smart cities. The cities are U-shaped models with high-level human capital infrastructures.

Kumar et. al. [12] proposed a machine learning-based approach for air quality analysis and prediction in the context of Indian smart cities. By observing the key variables of air pollutants, they implemented five machine learning models to predict Air Quality. The results of the models were compared with standard metrics and standard performance parameters. After evaluating the model, they found that GNB models performed better among all other models, and SVM models exhibited lower accuracy during prediction. The XGBoost model performed better than other models to get the linearity between actual and predicted values.

**2.2 Big Data for Smart city**

Zheng et al. proposed an approach to real-time air quality detection for any given location using environmental and historical air quality data. They designed a semi-supervised model that includes meteorology, human mobility, and city dynamics. Apart from that, they proposed an entropy minimization model for the best location prediction for monitoring air quality. The proposed approach is evaluated using the Beijing air quality dataset [13].

Lyu et al. presented a comparative study for cloud driven learning and big data technologies to solve the influencing factor for automation in a healthy smart city. It used some of the known algorithms to evaluate the model and retention rate which is around 23% more than other algorithms shows how useful it is to integrate such technologies [14].

Zhang et al.[15] proposed the idea of wireless-based environmental pollution analysis (WSN-EPA) to reduce the effect of pollution particles in smart cities. They continuously monitored the city's air quality levels and the movement of vehicle transportation. The pollutant, such as smoke gasses and other pollutants, was taken as an input and a pollution monitoring system developed.

Koch et al. proposed a terminology for the potential of newly generated data for smart cities in the context of sustainable development goals (SDGs). To fill the gaps and achieve the SDGs, existing tools and SDG indicators are important sources available. For the case study, they examined a district in Berlin and showed that local citizen participation and unofficial data are important from a smart city perspective [16].

Honarvar et al. [17] proposed a novel approach for air pollution monitoring using cost-effective strategies. A model for particulate matter (PM) prediction was developed, which consists of heterogeneous sources of urban data and a transfer learning model. The result of the proposed approach performs better compared to features collected from air pollution sensors. For evaluation, data from Aarhus, Denmark, was used.

Dwevedi et al. proposed a study based on the key factors that affect the living conditions in smart cities, including health, environment, and demography. Monitoring these factors continuously generates big data, hence requiring technological innovations for a good management plan. Additionally, the authors also emphasized the importance of environmental factors and their impacts on smart cities [18].

Zaree et al. proposed an approach to enhance the prediction and speed values of real-level air quality pollution and the effect of weather conditions using k-means clustering algorithms and big data mining tools for a city pulse project. The evaluation of the model shows that temperature, moisture, and wind speed are causes of the low pollution density. Apart from RMSE and MSE, the statistical measures show high efficiency and accuracy using the proposed model [19].

**2.3 Complex Event Processing for Smart City Applications.**

Simsek et al. [20] proposed a CEP model for automated extraction of rules from unlabeled IoT data. Since manual rule implementation through CEP is always hard due to the diverse and complex nature of dynamic data, they proposed a novel approach to extracting rules using deep learning methods. Various deep learning methods were compared for accurate rule extraction and their success rates. The proposed model was evaluated using an air quality dataset collected in Aarhus, Denmark..

Liu et al. [21] proposed an auto-extraction framework by combining deep learning and rule mining algorithms. The proposed work is based on a two-layer LSTM model with attention mechanisms. The approach is divided into two phases: in the first phase, anomaly data is filtered and labeled from IoT data, in second phase rule patterns are mined through decision tree mining algorithms for higher accuracy. The proposed work may assist in air pollution regulation, strategies and predictions.

**2.4 Air Quality in Context of Indian Smart Cities**

Ketu et al.[22] proposed a quantitative air quality prediction model using a novel approach of recursive feature elimination using Random Forest Regression for Indian smart cities. Seven popular models were compared with the proposed model and validated using well-known statistical metrics. The proposed model outperforms other existing models using predictions of AQI and Air pollution ($NO_X$) for a cleaner environment.

Chauhan et al. proposed a Convolutional Neural network (CNN)-based model for an air quality dataset to detect patterns of prediction. The entire approach is divided into two parts: (i) data analysis phases (ii) to test the proposed model using classified data on accurate models. The air pollutant data was used for the last five years, from 2015-2020. In addition, $PM_{2.5}$ and $PM_{1.0}$ peak levels of these metrics affect all of the Indian cities [23].

Samal et al. [24] proposed a deep learning-based model, the Convolutional LSTM-SDAE (CLS) model, to monitor particulate matter (PM) levels based on the correlation with meteorological factors. They applied imputation techniques to fill in missing values and a CNN-LSTM for finding hidden features. A Bidirectional Gatted Recurrent Unit was implemented as an encoder and decoder, which fine-tunes the model's prediction result. The model was validated using a specific location, which shows improved accuracy and outperforms other state-of-the art models.

Table 1: Comparison of existing literature works.

| Reference | Rule Formation | Types of Dataset | Query Processing | Premises | Model Used |
|---|---|---|---|---|---|
| Kortoçi et. al. 2022[10] | No | IoT | No | Indoor | Portable low cost sensor based model for monitoring. |
| Cui et. al.[11], 2022 | Yes | IoT | No | Outdoor | Inverted U shape based model. |
| Zheng et al.[13], 2015 | No | IoT | No | Outdoor | Neural network based fine grained model. |
| Simsek et. al.[20], 2021 | Yes | IoT | No | Outdoor | Rule extraction framework based model. |
| Ketu et al.[22], 2022 | Yes | IoT | No | Outdoor | Recursive feature elimination based random forest. |

| Chandra et. al. [25], 2022 | Yes | Geospatial | No | Outdoor | SSN Ontology based forest fire management. |
| Chandra et. al. [26], 2023 | Yes | Health Care | Yes | Outdoor | SWRL based vector borne disease diagnosis and treatment. |
| Kumar et. al. [27], 2023 | Yes | IoT | Yes | Indoor | Event processing based model |
| Mdhaffar et. al. [28], 2017 | Yes | Healthcare | Yes | Smart Health | Threshold based model for heart failure prediction. |
| Khazael et. al [29], 2023 | Yes | Geospatial | No | Outdoor | GeoT-REX based CEP model. |

## 3 Materials and Methodologies

In this section, we present the preliminary requirements for carrying out this research work. Consequently, the section is split up into various subsections named as: dataset description, Apache Kafka processing, RDF conversion using GraphDB, Siddhi CEP engine, and rule extraction using CPCB standard parameters. In subsection 3.1 a comprehensive discussion about the CPCB dataset is presented. In subsection 3.2, Dataset preprocessing discussed in the context of the proposed work. Subsection 3.1.1 explores the proposed architecture and its working in detail. Subsection 3.2 shows the event driven stream processing using Kafka and Jena. Subsection 3.2.1 shows conversion of sensor data to RDF. In subsection 3.3 elaborates the siddhi CEP engine and its working for RDF data. Subsection 3.4 discusses the rule generation through using decision trees and CPCB standard parameters. Subsection 3.5 shows conversion of siddhi query language to SPARQL query. Subsection 3.6 shows experimental flow of proposed work.

### 3.1 Dataset description

For this research work, we have collected Air Quality data from the official website[3] of the CPCB which works under the aegis of the Ministry of Environment, Forests, and Climate Change, Government of India. The CPCB dataset includes various relevant factors such as ($PM_{10}$), fine Particulate Matter ($PM_{2.5}$), Nitrogen Monoxide (NO), Nitrogen Oxide (NOx), Nitrogen Dioxide ($NO_2$), Ammonia ($NH_3$), Sulphur Dioxide ($SO_2$), Carbon Monoxide (CO), and Ozone ($O_3$) and its corresponding location along with respective AQI bucket.

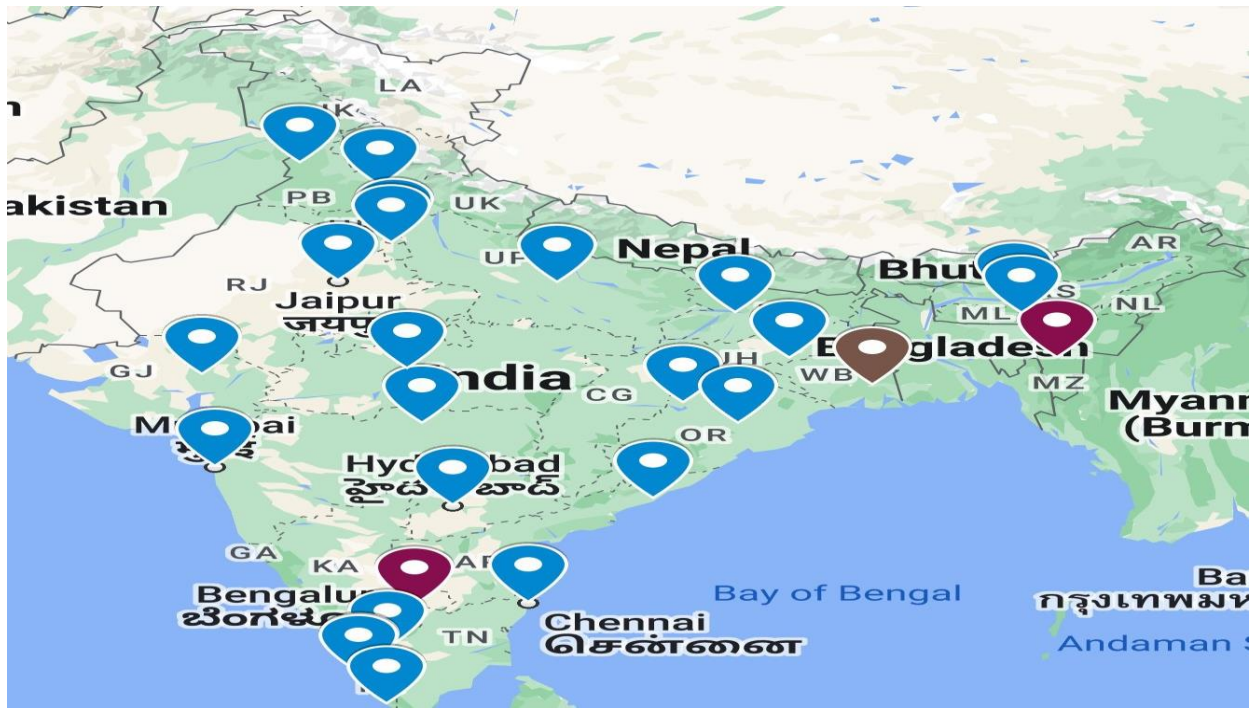

Fig 1: Different base locations for Air Quality of Dataset collection

The spatial dataset contains (25932 rows and 16 columns) samples collected from January 1, 2015, to January 1, 2020. These air pollutants play an important role in monitoring the air quality of location in real time. The fig.1 shows the various locations of the dataset collected for monitoring purposes.

**3.2 Dataset preprocessing**

This step in research work is important due to the significance of the dataset. During the IoT dataset collection, some of the important missing values may change the efficiency of the proposed work. Since the dataset contains numerous air pollutant attributes and each of these has some missing values available, we have used the median value to fill in the empty values.

---
[3] https://www.kaggle.com/datasets/rohanrao/air-quality-data-in-india

Fig. 2 and 3 show the dataset before and after preprocessing. The median of that particular attribute fills in the missing value.

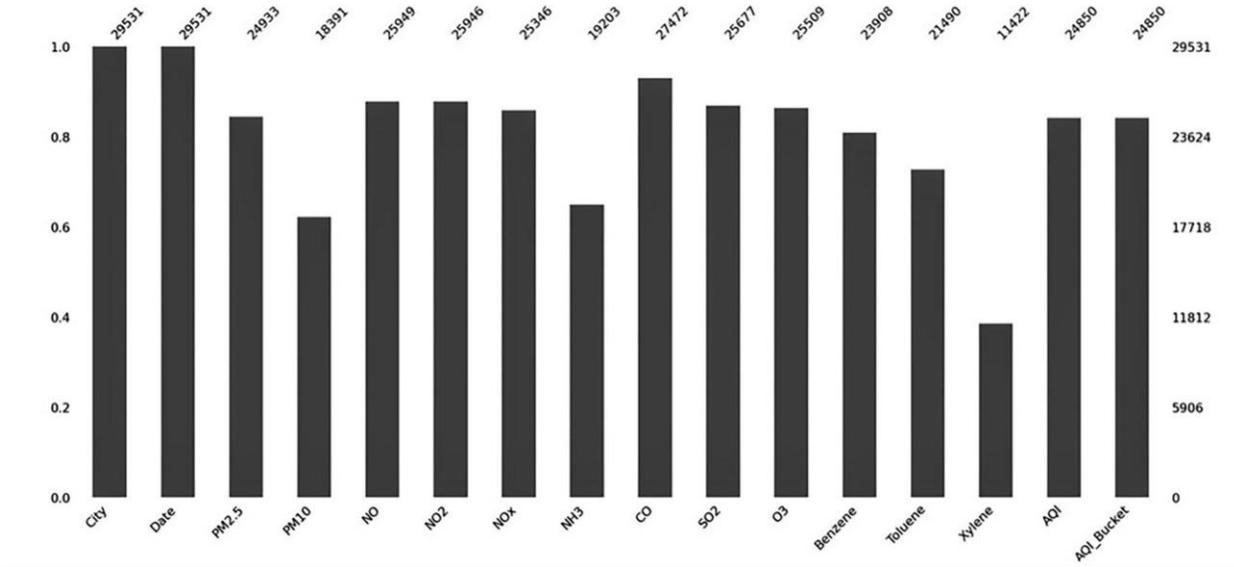
Fig 2 : Dataset before preprocessing

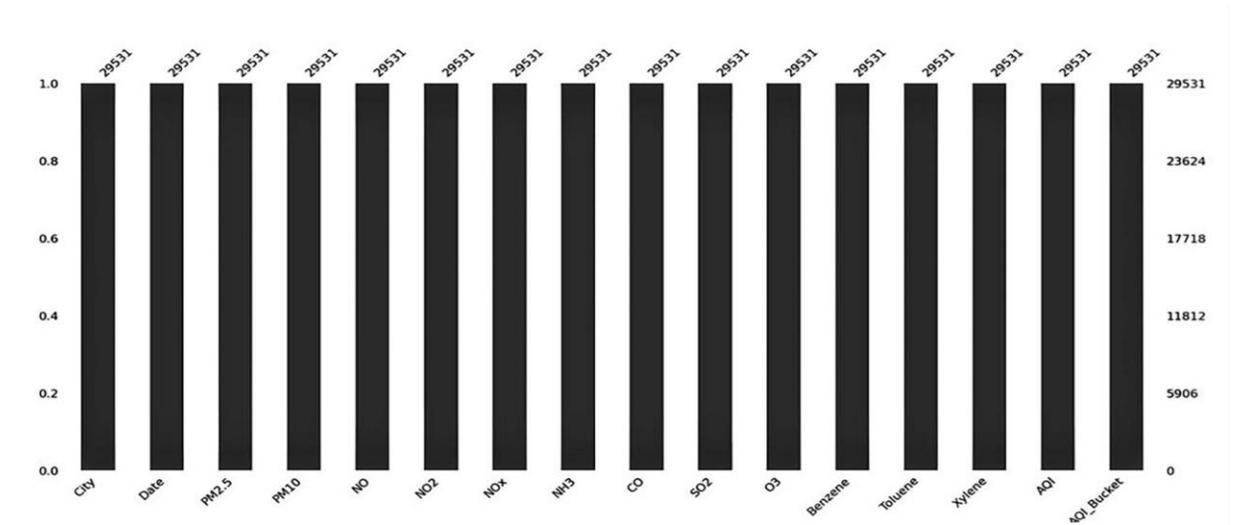
Fig 3 : Dataset after preprocessing

Table 2:Description of the dataset

| Air pollutant | Measurement unit | Duration of data collection | Data type | Source |
|---|---|---|---|---|
| $SO_2$ | Micrograms per cubic meter | January 1, 2015, to January 1, 2020 | Time Series | CPCB |
| NO | Micrograms per cubic meter | January 1, 2015, to January 1, 2020 | Time Series | CPCB |
| $NO_x$ | Micrograms per cubic meter | January 1, 2015, to January 1, 2020 | Time Series | CPCB |
| CO | Micrograms per cubic meter | January 1, 2015, to January 1, 2020 | Time Series | CPCB |
| $NH_3$ | Micrograms per cubic meter | January 1, 2015, to January 1, 2020 | Time Series | CPCB |
| $O_3$ | Micrograms per cubic meter | January 1, 2015, to January 1, 2020 | Time Series | CPCB |
| $PM_{2.5}$ | Particulate Per Matter | January 1, 2015, to January 1, 2020 | Time Series | CPCB |
| $PM_{10}$ | Particulate Per Matter | January 1, 2015, to January 1, 2020 | Time Series | CPCB |
| AQI | Specific range values | January 1, 2015, to January 1, 2020 | Time Series | CPCB |

The raw dataset is preprocessed to remove irrelevant and missing data and make it suitable for further analysis. Table 2 presents summaries of the attributes of the dataset, measurement unit, duration of the dataset collected, data types, and sources of the dataset. These multi-parameters of datasets play an important role in further extracting information.

### 3.1.1 Proposed Architecture

Fig. 4 shows the proposed architecture for smart city-based air quality monitoring. Initially, the dataset is collected from the CPCB open repository for preprocessing, and RDF conversion is performed using the GraphDB. We have developed the Knowledge Graph based on CPCB air quality guidelines in the context of Indian cities and streamed the RDF data using Apache Jena. The aim is to ensure that efficient query processing for smart city monitoring is performed based on the rules formed using CPCB standard parameters.

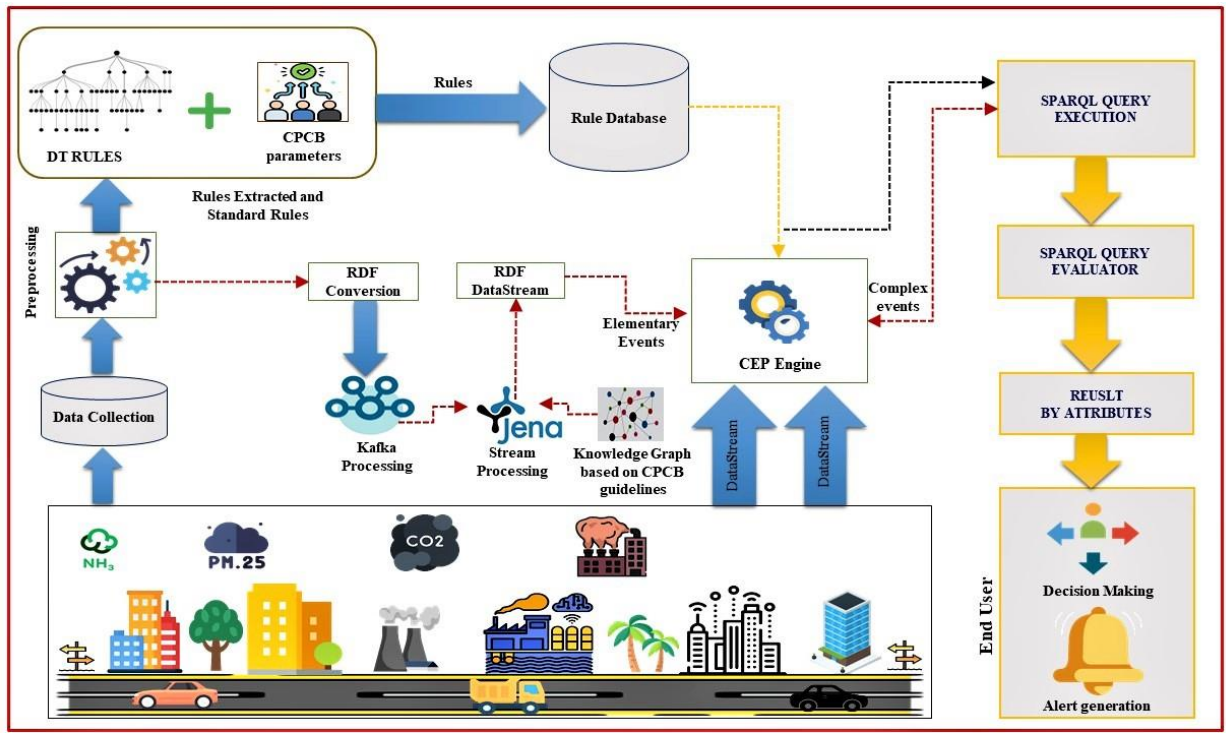

Fig. 4 :Proposed architecture for Rule based Complex Event Processing for Air Quality Monitoring (RCEPAQM) .

### 3.2 Event Driven Stream Processing

In the Big Data environment, stream processing is an important component for realizing the data in real time. The smart city generates a bulk of data that needs to be processed simultaneously to make an efficient decision-making process that will be helpful in uplifting the living standards of residents in every aspect. In this research, we are leveraging the services of Apache Jena and Apache Kafka [30], which can handle millions of events very efficiently. Kafka is based on stream processing, which has four key concepts named: (i) broker (ii) topics (iii) producer (iv) consumers [31].

The Kafka topic provides a way of organizing that serves as a data container for the immediate processing of data between applications and systems. Each topic is defined by partitioning for faster data processing as well as data redundancy. Kafka producers write about the topics, and later they are consumed by consumers. The broker is a server that manages the actual reading, writing, and load balancing.

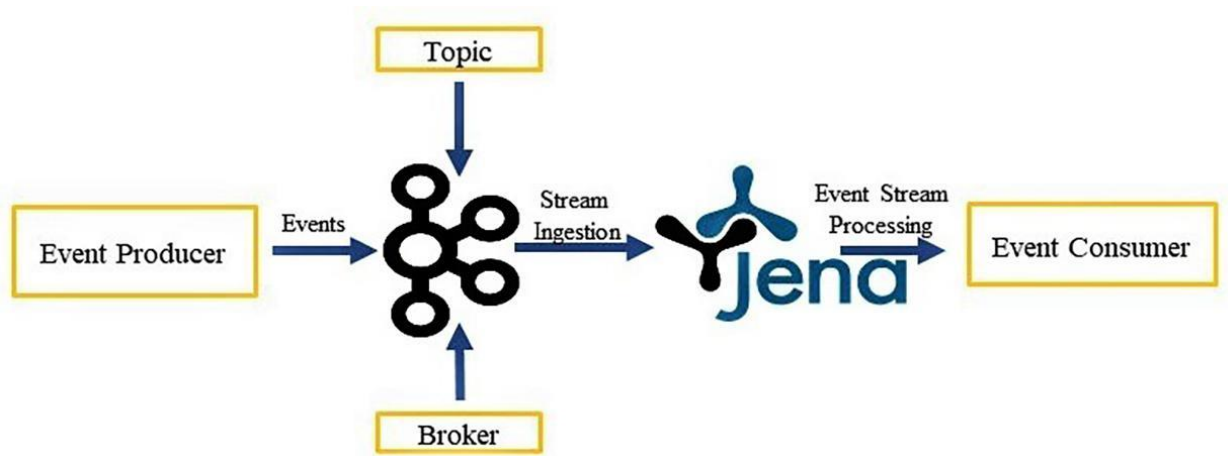

Fig 5: An integration of Apache Kafka with Apache Jena

The producer of Apache Kafka produces the events for Kafka clusters. Furthermore, the Kafka connectors allow you to configure sources and sinks that link Kafka topics to well-known applications or data systems using common interfaces such as JDBC (relational SQL databases), Hadoop Distributed FileSystem (HDFS), and Amazon S3. Custom connectors are used to connect the CEP to the data layer and knowledge discovery layer, automating the transmission of information between the CEP and both levels [32].

Apache Jena is an open-source, Java-based framework for building semantic web applications. It employs powerful tools and APIs (Application Programming Interfaces) for real-time SPARQL query processing. Jena utilizes reasoning-based rule engines to handle complex tasks and draw useful conclusions. Fig. 5 shows the integration of Apache Jena with Apache Kafka for real time processing.

**3.2.1 Sensor Data to RDF Conversion**

The Air Quality data collected from the CPCB of India is formatted into CSV format, and converted into RDF data using the RDFlib[4], which is a python library available with compliance with the W3C standard. The RDF provides the ability for interoperability and integration of data across sources, made possible by the RDF capacity to establish linkages across resources. This is especially beneficial for programmes that need to combine data from several sources or areas.

After conversion of CSV into RDF, it forms a triplet form named Subject, Predicate, and Object Format, where RDF [25], [26] expands the Web's linking structure by using URIs to specify the connection between objects as well as the two ends of the link. Structured and semi-structured data may be combined, exposed, and shared across several applications using this straightforward architecture.

### 3.2.2 Knowledge Graph Development

For Knowledge Graph Development we have used Apache Jena framework and followed the guidelines of CPCB air quality monitoring. Afterwards, integrated the RDF data to a knowledge graph which gives the precautionary suggestion that air quality differs from city to city. According to the situation we extracted information through a query on knowledge which contains complet RDF data.

### 3.3 Siddhi CEP

Siddhi is a CEP engine that works with a stream of events to analyze and correlate them to extract useful information patterns for end user decision-making. Building event-driven applications for use cases like real-time analytics, data integration, notification management, and adaptive decision-making requires a fully open source, cloud-native, scalable, microstreaming, and CEP system[5]. Fig. 6 shows an architecture of event processing through Siddhi CEP engine.

---

[4] https://rdflib.readthedocs.io/en/stable/

[5] https://siddhi.io/

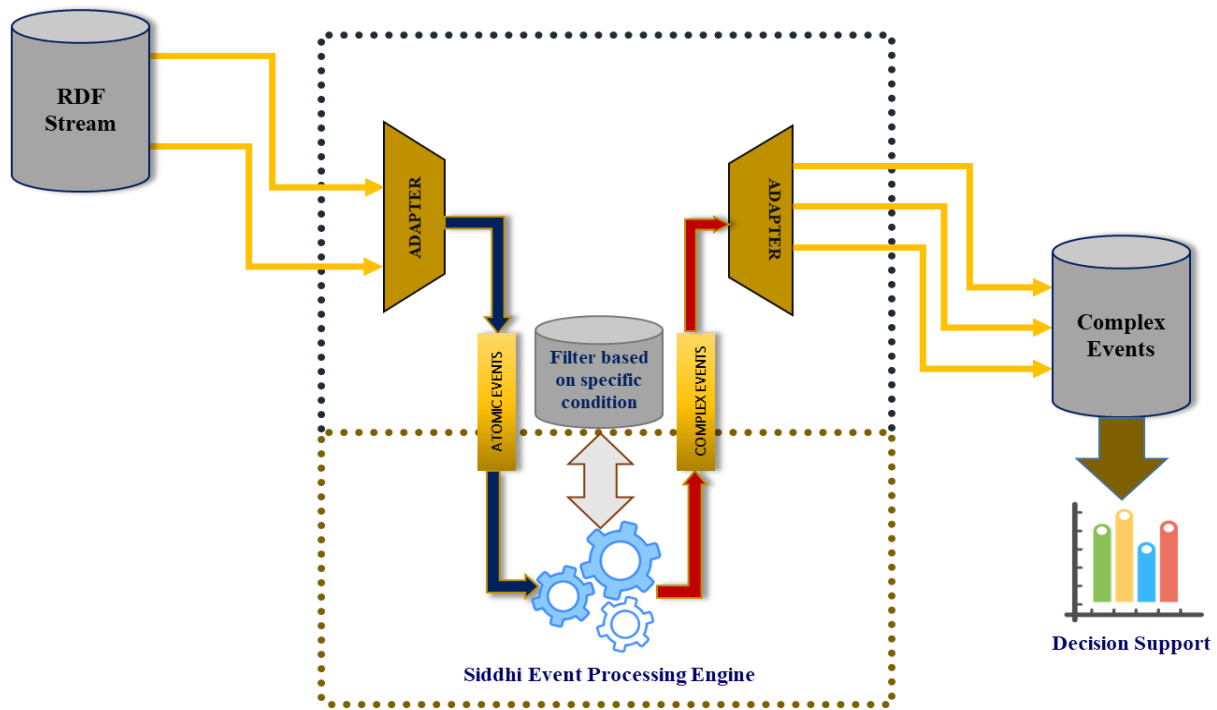

Fig 6: An architecture of Siddhi CEP

In this research work, the Siddhi CEP engine takes input in the form of air quality data and finds a useful pattern based on the specific rules and conditions extracted through certified parameters. The Siddhi CEP engine translates the specified and analyzed rules into CEP queries. The queries are processed using the SPARQL query with the help of the GraphDB tool for static data, and for stream data, we have used Apache Jena.

### 3.4 Rule Development

This subsection emphasizes the rule generation phase of this work. These rules are developed using decision tree algorithms and standard metrics for each parameter as proposed in CPCB guidelines[6]. For each of the air pollutants, specific values are prespecified, and it helps to monitor the values for each parameter, and once they are exceeding or decreasing, an appropriate decision may be taken for end users.

---

[6] https://cpcb.nic.in/air-quality-standard/

We have developed certain criteria that are effective for categorizing the air quality based on the AQI, which are based on the National Ambient Air Quality Standards based on CPCB guidelines and decision tree. Some of the rules created shown in table 2.

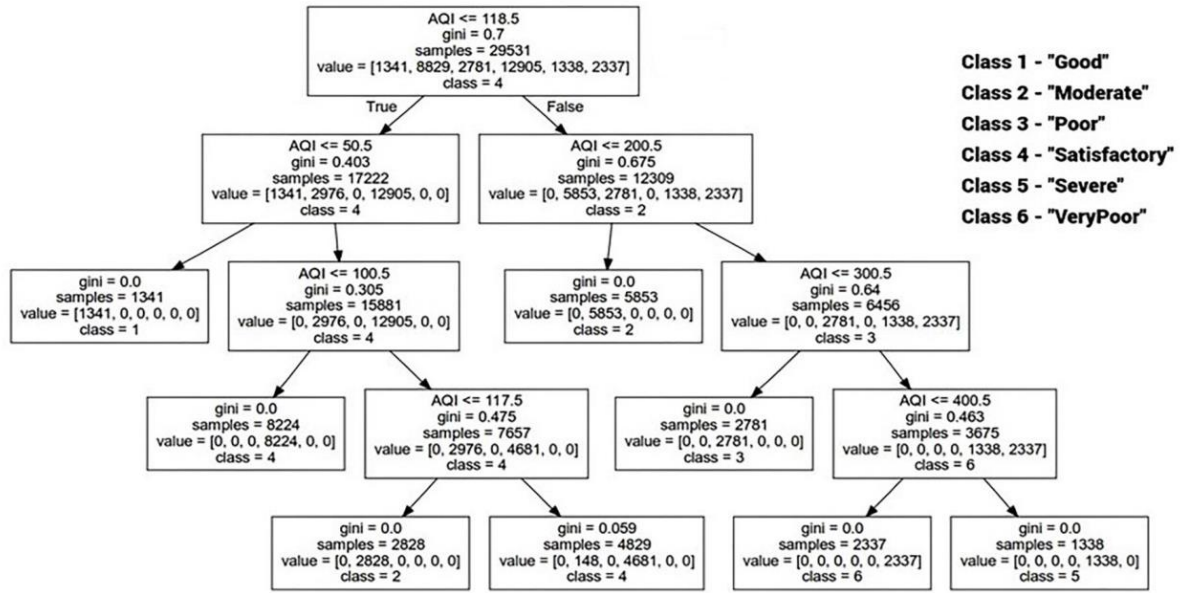

Fig.7 : Decision tree formed for rule generation.

Fig. 7 shows the decision tree created for rule generation. Pruning has been used for an optimized size of the decision tree, which may reduce the chance of the model overfitting.

Table 3: Rules formed using CPCB standards and decision tree.

| S.NO | Rules | AQI(Category) |
|---|---|---|
| 1 | If(PM2.5 >= 55.5 and PM2.5 <= 150.4) and ($O_3$ >= 51 and $O_3$ <= 75) | Moderate |
| 2 | If(PM10 >= 255 and PM10 <= 354) and ($O_3$ >= 70 and $O_3$ <= 101) and (NOx >= 6) | Poor |
| 3 | If (PM2.5 >= 250.5 and PM2.5 <= 50) and ($SO_2$ > 20 and $NO_2$ > 60 and $NO_2$ < 140) | Severe |
| 4 | If(PM10 >= 325 and PM10 <= 600) and ($SO_2$ > 20 and NOx > 60) and (NOx < 160 and NO > 100) and $NH_3$ < 100 | Severe |
| 5 | If (PM2.5 >= 0 and PM2.5 <= 12) and ($NO_2$ >= 8 and $NO_2$ <= 18) and ($NH_3$ >= 10 and $NH_3$ < 17) | Good |

The table 3 represents the AQI and its ranges to classify the category as per the values. Each of the air pollutants has a specific value, and based on the concentration range rules, they have been designed. These values have been given by the CPCB[7], an authorized body for taking control of air pollution in India. Using the decision tree and CPCB standard we have designed thirty rules and executed various queries.

The following steps show how AQI is calculated based on the values of air pollutants.

1. The sub-indices for each pollutant at a monitoring site are created using the 24-hour average concentration value (8-hourly for CO and $O_3$) and the health breakpoint concentration range. The AQI is the worst sub-index for that region.
2. Not all facilities might be able to monitor each of the eight contaminants. The total AQI cannot be calculated unless data for at least three pollutants, one of which must be either PM2.5 or PM10, are available. In any other circumstance, the information would not be judged adequate to calculate the AQI. Similar to the subindex, it is believed that at least 16 hours of data are needed to compute it.
3. The sub-indices for identified pollutants are calculated and reported even when there is not enough data to calculate the AQI. The individual pollutant-wise sub-index provides the air quality status for the given pollutant.

Table 4 : AQI level

| AQI Category | AQI | Concentration Range | | | | | | | |
|---|---|---|---|---|---|---|---|---|---|
| | | $PM_{10}$ | $PM_{2.5}$ | $NO_2$ | $O_3$ | CO | $SO_4$ | $NH_3$ | PB |
| **Good** | 0-50 | 0-50 | 0-30 | 0-40 | 0-50 | 0.0-1.0 | 0-40 | 0-200 | 0-0.5 |
| **Satisfactory** | 51-100 | 51-100 | 31-60 | 41-80 | 51-100 | 1.1-2.0 | 41-80 | 201-400 | 0.5-1.0 |
| **Moderately Polluted** | 101-200 | 101-250 | 61-90 | 81-180 | 101-168 | 2.1-10 | 81-380 | 401-800 | 1.1-2.0 |

---

[7] https://app.cpcbccr.com/AQI_India_Iframe/

Table 4 shows the AQI standard guidelines and various air pollutants compositions according to CPCB.

**3.5 Conversion for rules of SiddhiQL queries into SPARQL queries.**

Conversion of Siddhi Query Language (SiddhiQL) queries into RDF-based SPARQL queries requires a unique API or script to carry out the conversion when converting SiddhiQL queries into SPARQL queries programmatically. In this procedure, SiddhiQL queries are parsed in order to produce SPARQL queries that are appropriate for the data model and needs. Applications for the semantic web and linked data frequently employ SPARQL Interfacing Notation [33] to improve the querying and reasoning power of RDF data. It is especially useful in situations where data has to be enhanced, verified, or changed in accordance with specified guidelines and specifications. To carry out inferencing and data management activities, SPIN may be utilized with RDF data stores and graph databases.

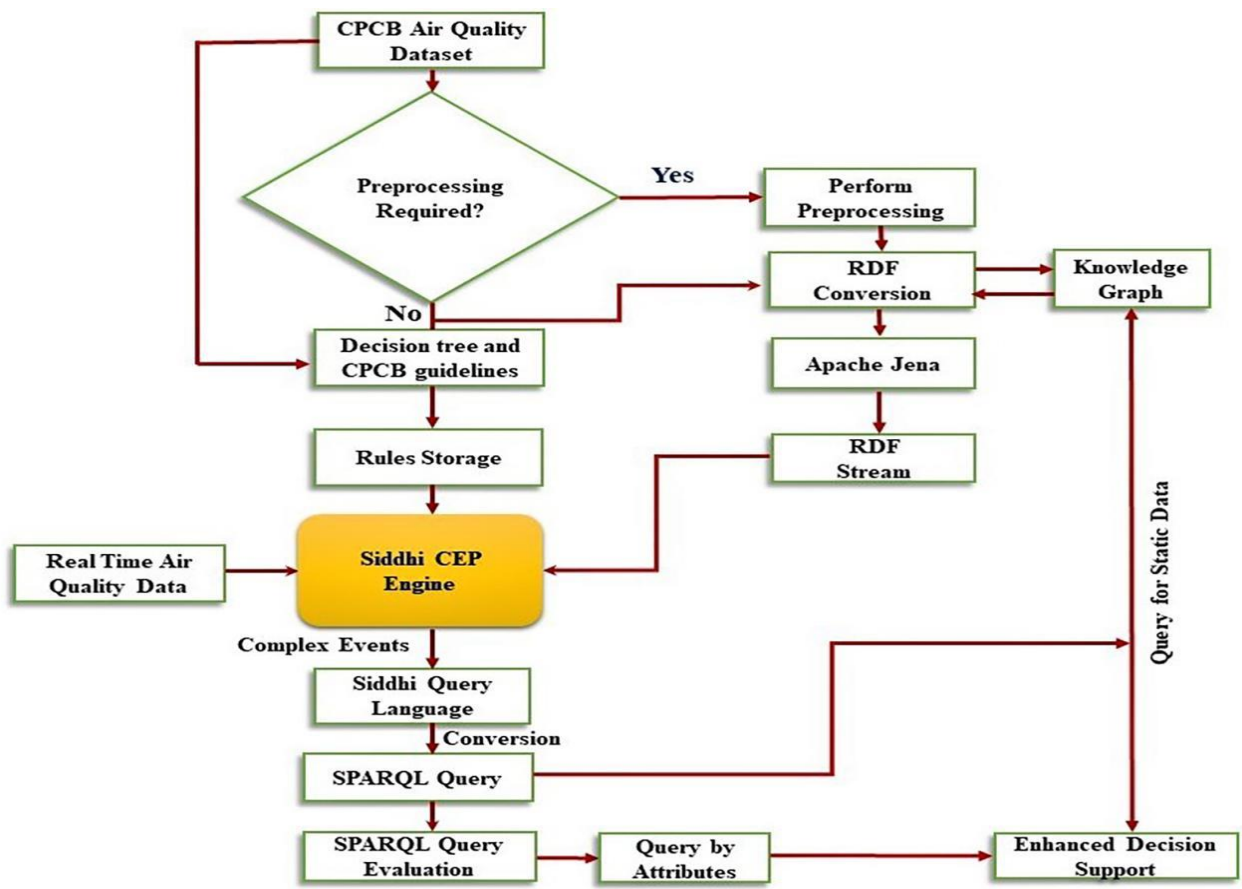

Fig.8 :-Experimental work flow of the proposed work

Fig. 8 shows the experimental workflow of the entire proposed work and how it flows through RCEPAQM.

## 3.6 Use Case of Proposed Work

In this subsection, a smart city-based use case is discussed using the proposed framework, as shown in Fig. 4.

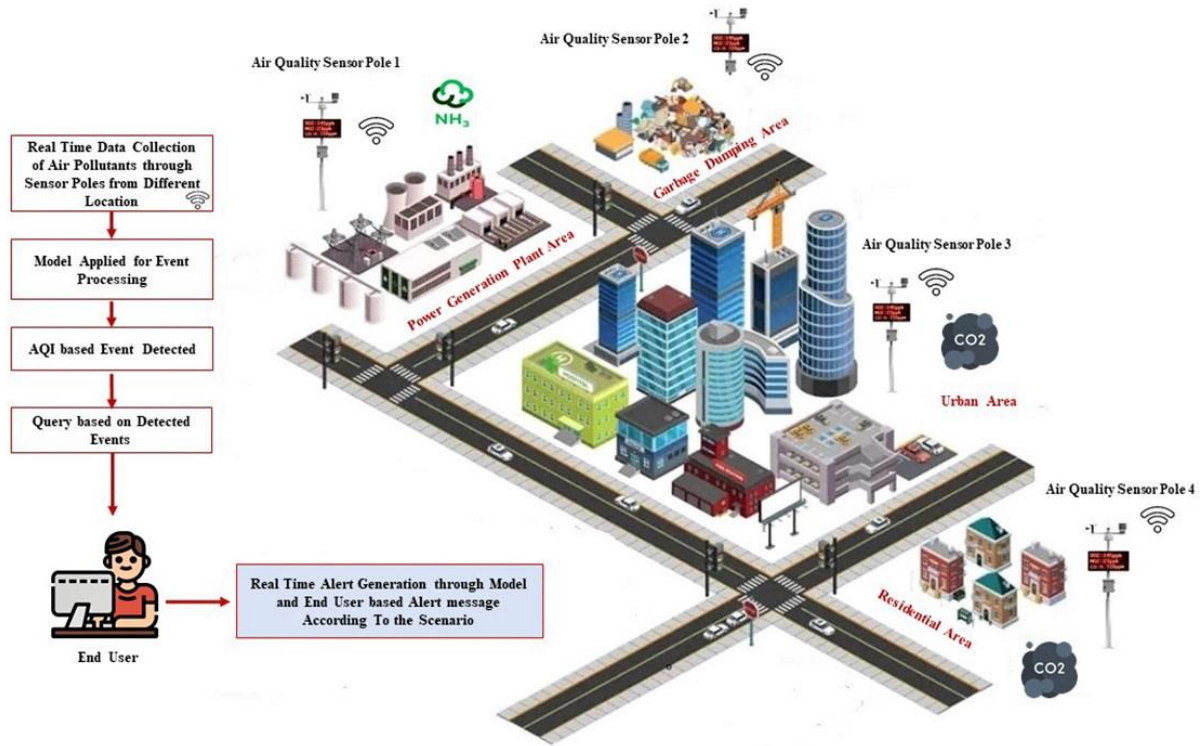

Fig.9 :- Use case scenario of Air Quality Monitoring for cities

According to Fig. 9, there are many AQI sensor poles installed at different locations in smart cities that collect air quality details and send them to the air quality control center in real time. Pole 1 is located in the power generation plant area; pole 2 is located in the garbage dumping area; pole 3 is located in the urban area of the city; and pole 4 is located in the residence area. The most affected areas are pole 1 and pole 2, as shown in Fig. 9, because they emit more harmful gases that affect the AQI. When these gases reach poles 3 and 4, the area also gets affected. According to this situation, our model works well to give real-time alerts based on the collected real-time data on data centers. The model applied gives alerts based on events and sends the alerts to the stakeholders. The knowledge graph provides additional information based on alerts and precautionary measures of CPCB guidelines to the stakeholders.

**4 Result and Discussion**

According to this paper, RCEPAQM is a novel approach to measuring air quality in Indian cities using rule-based CEP [34] [35]. For monitoring purposes, we used the CPCB dataset and transformed it into RDF data so that it could be used for query and event processing. A knowledge graph is added to track the AQI and find the important pollutants for determining whether they are within the permitted ranges or not based on rules created using CPCB standard parameters and a decision tree. Simultaneously, the RDF data is split into various chunks for optimal query processing execution time, and various types of queries are tested on different RDF data.

Furthermore, the Siddhi CEP engine is used for event processing, and a huge number of events are sent to observe the amount of time it takes to execute events based on rules created. In addition, different RDF chunks are used to keep track of event processing time for improved monitoring.

The experimental process has been covered in various phases, respectively, as follows: (1) In this phase, we demonstrate the performance of the Siddhi CEP engine in terms of query deployment and processing time. (2) We evaluate the proposed work with different chunks of the RDF dataset. (3) Evaluation metrics of events tested and deployed based on the rules. To evaluate the performance of the proposed event processing model, our system is configured with an Intel(R) Core(TM) i7-7700 CPU @ 3.60 GHz, 16 GB of RAM, a 64-bit operating system, an x64-based processor, four cores, and the Linux Operating System.

Mdhaffar et al. [36] proposed a novel approach to event processing for heart failure prediction using the MIMIC II waveform dataset. The model was tested with different types of queries, named simple filter query and pattern matching query. This work also resembles a similar type of processing with a number of events in the context of AQI for air quality monitoring. While comparing both, our work resembles a better perspective on performance while increasing the number of events.

4.1 **Different SPARQL results**

In this subsection various queries are executed using Apache Jena and GraphDB. Fig. 10 and 11 shows the execution of Q1 and Q2 through Apache Jena while Fig. 12 and 13 shows execution of queries through various GraphDB.

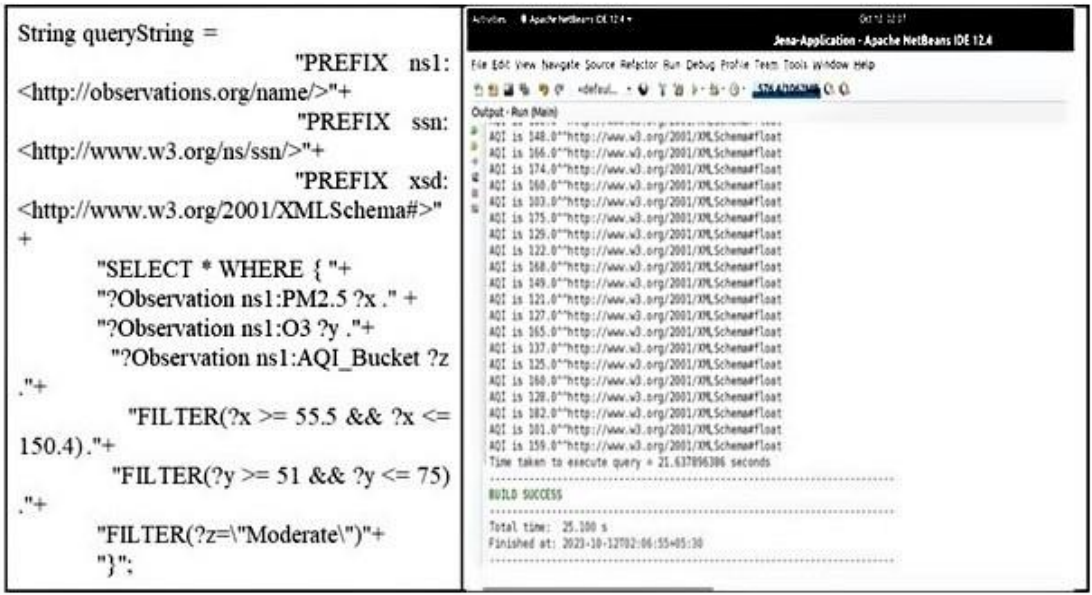

Fig. 10 Execution of Q1 based on batch processing

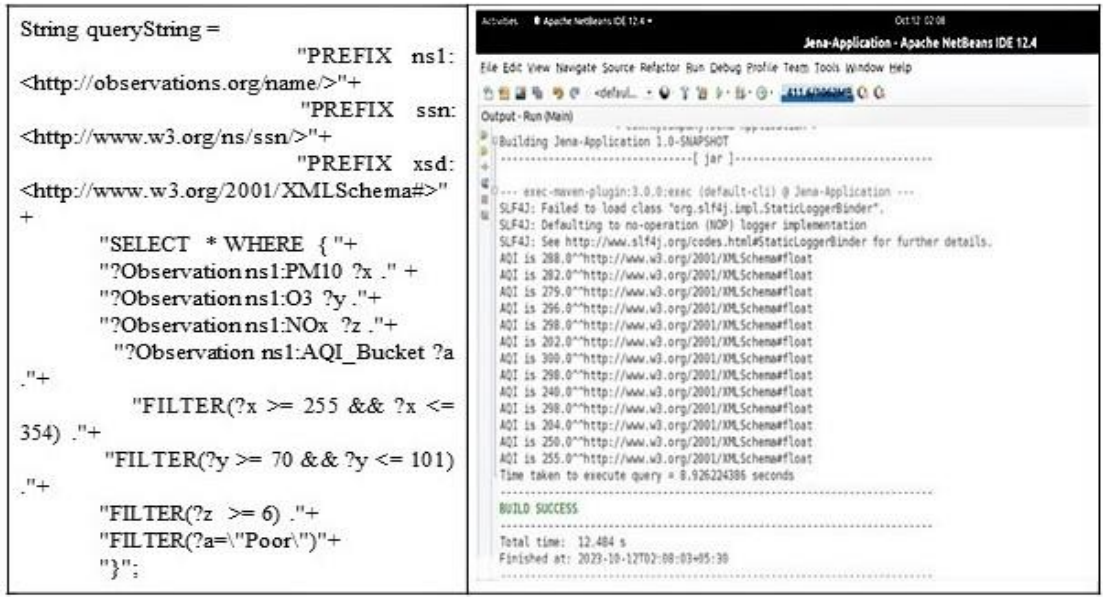

Fig. 11 : Execution of Q2 based on batch processing

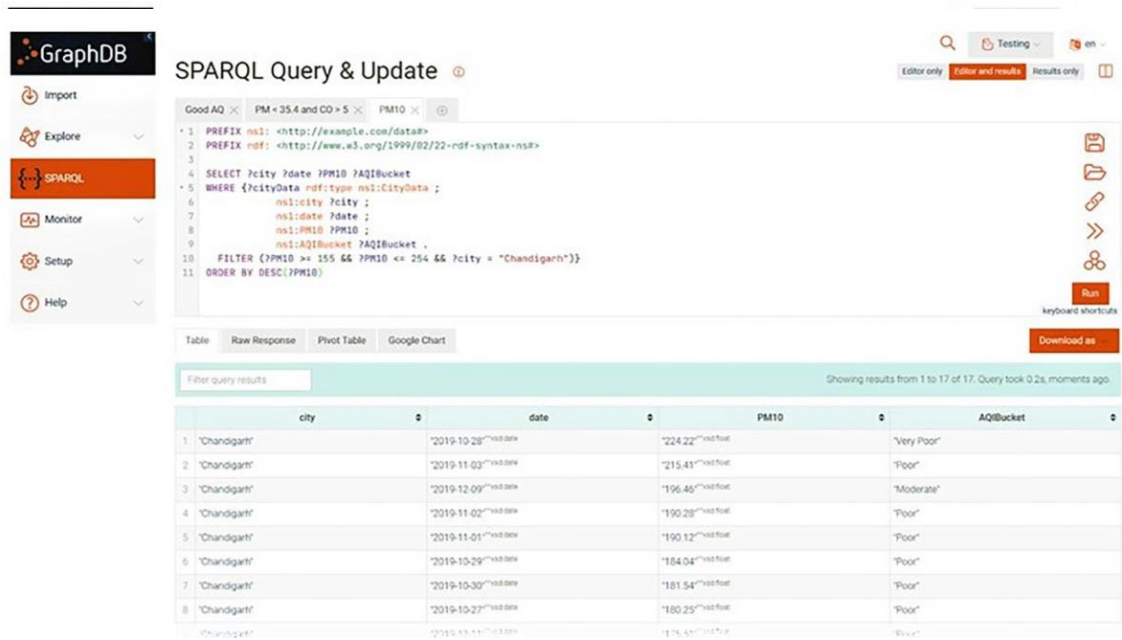

Fig. 12 Query result through GraphDB

## 4.2 Event Processing Time

We have implemented and compared several queries to the Siddhi CEP engine and observed the events based on the criteria developed for categorizing the air quality dataset according to AQI. To evaluate the performance measure, we deployed simple as well as complex queries and observed the time taken based on a large number of events. As indicated in Fig. 8 below, the event execution time in seconds is gradually increasing. It is evident that a simple query takes much less time to process than complex queries. We processed an enormous number of events through the CEP engine and observed that the execution time is varying as the number of events increases. Table 5 shows the events execution time through Siddhi CEP engine.

Table 5: Events execution time through Siddhi CEP Engine

| Number. of events | Rule 1 | Rule 2 | Rule 3 | Rule 4 | Rule 5 |
| --- | --- | --- | --- | --- | --- |
| 5000 | 2.95s | 2.96s | 2.95s | 2.94s | 2.94s |
| 10000 | 2.98s | 2.95s | 2.94s | 2.88s | 2.94s |
| 15000 | 2.96s | 2.97s | 2.94s | 2.92s | 3.08s |

Fig. 14 shows the rule processing when the number increases and its execution time in seconds.

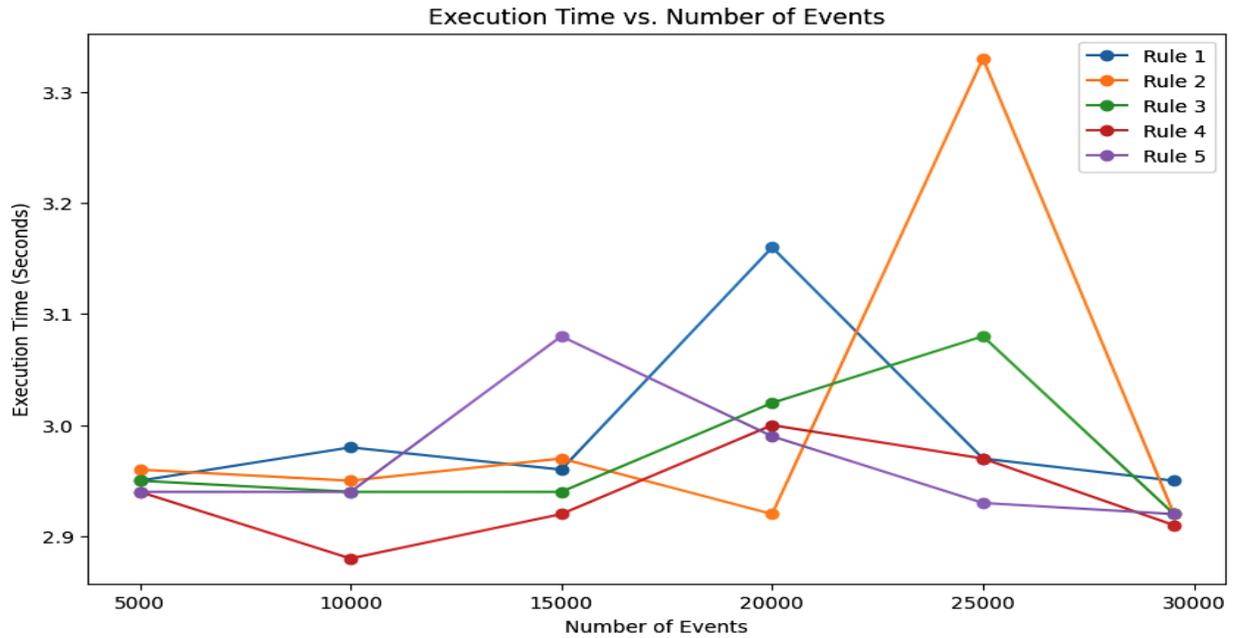

Fig.14 : Rule processing time as events rise in number

### 4.3 Execution Time for CEP rule deployment

Through the experiments, we tested the proposed model for the updated rule insertion time based on a CEP engine after updating into the filtered city data, which is the actual datastream. Rule deployment time is evaluated based on various parameters and pattern matching through simple as well as complex queries. This work deploys the rules based on query execution on a data stream. We have tested 1000 to 8000 events in CEP and observed the execution time. We found that the time is increasing when the number of events is increasing. Table 6 shows the rule deployment time while executing events.

Table 6 : Rule Deployment Time in Seconds

| Number of Events | 1000 | 2000 | 3000 | 4000 | 5000 | 6000 | 7000 | 8000 |
|---|---|---|---|---|---|---|---|---|
| Rule Processing Time | 4.12 | 5.23 | 6.01 | 7.29 | 8.11 | 9.14 | 10.25 | 11.37 |

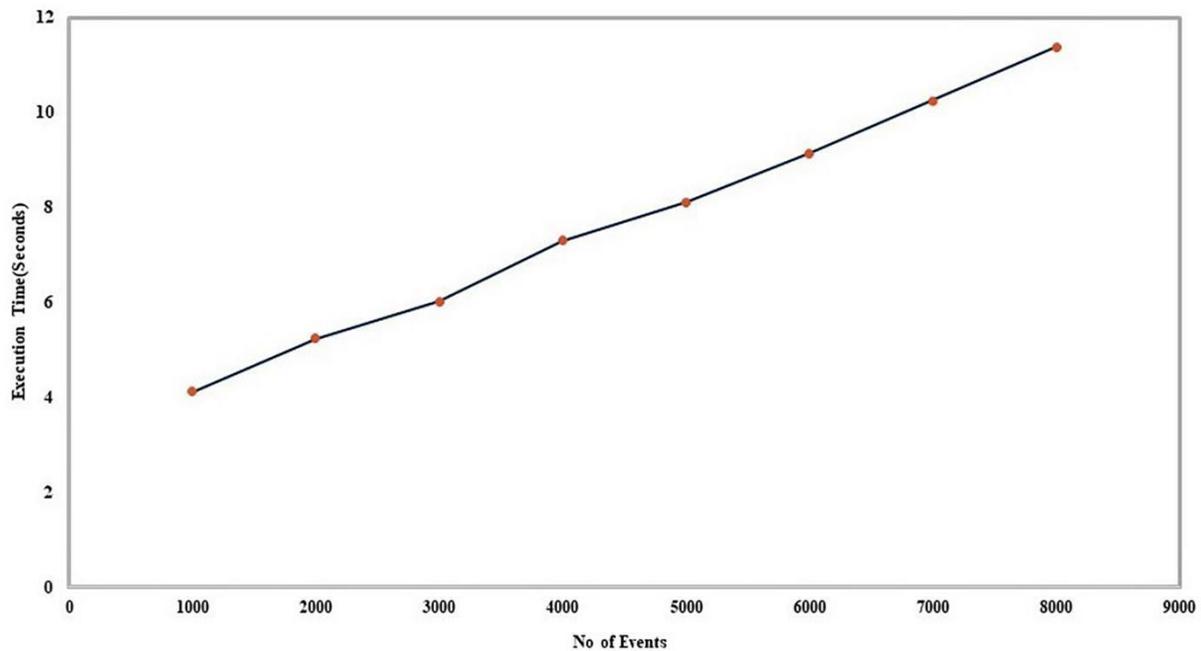

Fig.15 : Rule processing time as events rise in number

From Fig. 15, it is clearly visible that the number of events increases while rule processing time also increases. The rule processing time is measured in seconds.

### 4.3 Query Processing based on Different RDF Chunks

This subsection elaborates on the query processing time based on different chunks of the RDF dataset. We have divided the RDF dataset into different chunks named Chunks A, B, C, D, and E, which corresponds to RDF 1, RDF 2, RDF 3, RDF 4, and RDF 5, and based on that, we process the queries and observe the execution time. Since we have two types of queries: simple and complex, based on evaluation with various RDF datasets, we observed that the query processing time varies based on the events of the RDF chunks. Similarly, we have considered various combinations of RDF chunks, such as two, three, and four at a time, to execute similar queries. We have shown five queries for each type of RDF combination.

Table 7: Query execution time using various RDF chunks without combination

| Query Numbers | Chunk-A | Chunk-B | Chunk-C | Chunk-D | Chunk-E |
|---|---|---|---|---|---|
| Q1* | 1.13s | 1.58s | 1.81s | 1.27s | 1.25s |
| Q2 | 0.68s | 1.02s | 0.78s | 0.87s | 0.78s |
| Q3 | 0.79s | 1.03s | 0.96s | 0.98s | 0.82s |

Table 7 shows the various query execution times (in seconds) where query Q1, Q2, Q3, Q4, and Q5 have been executed on individual RDFs. The Q1* is complex and all others are simple queries. Table 8 shows the combination of two different RDFs and their impact on queries execution time.

| Query Numbers | Chunk-A & B | Chunk-A & C | Chunk-A & D | Chunk-A & E | Chunk-B &C |
|---|---|---|---|---|---|
| Q1* | 3.99s | 4.52s | 3.72s | 4.12s | 4.30s |
| Q2 | 1.70s | 1.70s | 1.62s | 1.44s | 1.84s |
| Q3 | 1.46s | 1.40s | 1.54s | 1.49s | 1.76s |

Table 8: Query execution time using two RDF chunks

Table 9: Query execution time using three RDF chunks

| Query Numbers | Chunks-A&B&C | Chunks-A&B&D | Chunks-A &B & E | Chunks-A&C&D | Chunks-A&C&E |
|---|---|---|---|---|---|
| Q1* | 5.88s | 5.43s | 5.37s | 5.81s | 5.23s |
| Q2 | 2.67s | 2.67s | 2.74s | 2.76s | 2.50s |
| Q3 | 2.47s | 2.61s | 2.44s | 2.75s | 2.69s |
| Q4 | 2.57s | 2.59s | 2.26s | 2.39s | 2.73s |
| Q5 | 2.83s | 2.85s | 2.91s | 2.83s | 3.25s |

Table 9 shows the various query execution times based on combining three RDFs at a time. The query execution time differs from other scenarios for similar types of queries. It is observed that for each of the queries, various RDFs combinations have different execution times. Table 10 shows the combination of four RDFs and its performance for each query execution.

Table 10: Query execution time using four RDF chunks

| Query Numbers | Chunks-A&B C&D | Chunks-A & B&C&E | Chunks-A& B& D&E | Chunks-A & C & D & E | Chunks-B & C & D &E |
|---|---|---|---|---|---|

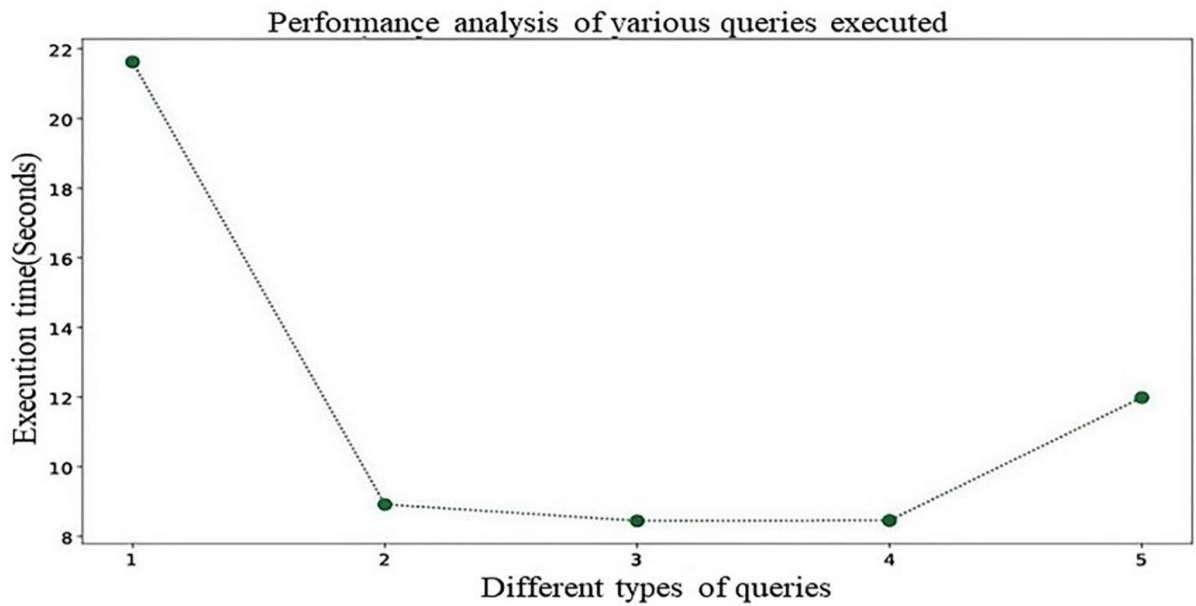

| Q1* | 6.78s | 7.79s | 6.59s | 6.61s | 6.72s |
| --- | --- | --- | --- | --- | --- |
| Q2 | 2.90s | 2.78s | 2.84s | 3.10s | 2.89s |
| Q3 | 2.91s | 2.42s | 3.07s | 3.56s | 2.76s |
| Q4 | 2.95s | 2.66s | 3.44s | 3.30s | 2.88s |
| Q5 | 3.56s | 3.12s | 3.57s | 3.86s | 3.41s |

Fig.16 : Execution analysis of various queries on entire RDF data Fig. 16 shows the execution time of different queries on the entire RDF data. From Tables 5, 6, 7, and 8

it is observed that query execution time varies by using two, three, and four combinations of RDF data.

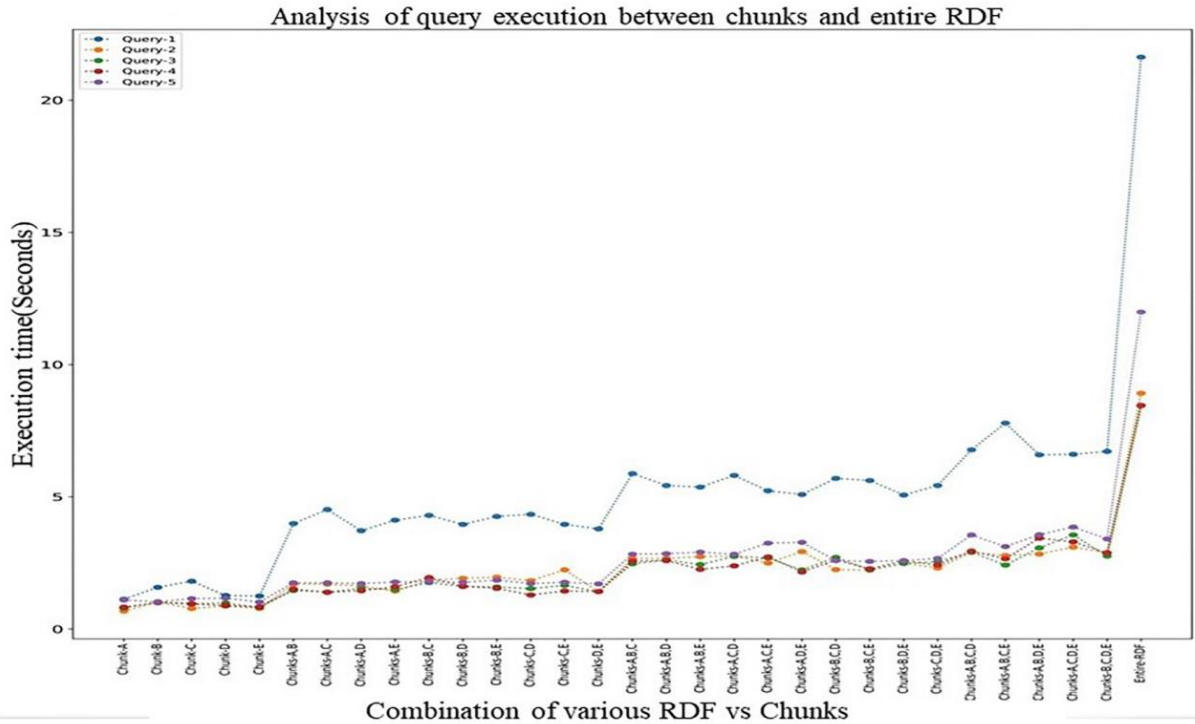

Fig.17 : Comparative analysis of query execution on combination of different RDFs chunks vs combination of entire RDF

Table 11: Queries performance analysis for static and stream data

| Queries | Static RDF | RDF Stream(Batch) |
|---|---|---|
| Q1* | 44.27s | 21.63s |
| Q2 | 16.36s | 8.92s |
| Q3 | 23.58s | 8.45s |
| Q4 | 27.11s | 8.46s |
| Q5 | 31.33s | 11.99s |

Fig. 17 shows the comparative analysis of different RDF chunks vs entire RDF based query execution. Table 11 shows the various query execution times (in sec) using static RDF data and RDF data streams. A total of five queries (Q1*, Q2, Q3, Q4, and Q5) have been used for comparison, where Q1* is a complex query and others are simple queries. We observe that the SPARQL query execution time on static RDF is higher as compared to the datastream.

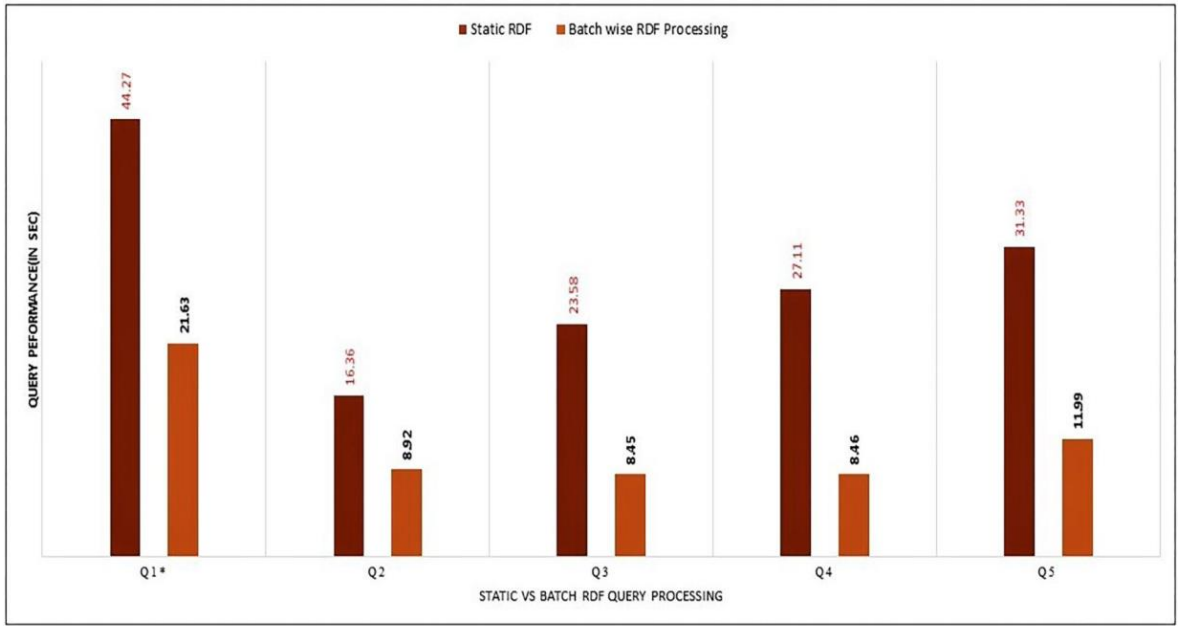

Fig.18 : Query performance comparison for static RDF vs RDF stream

Fig. 18 and 19 shows the query performance time using static RDF vs. RDF stream and RDF chunk-based event execution time. Table 12 shows performance of RDF chunks based on event execution time.

Table 12: Performance of event processing on various RDF data.

| RDF chunks | Number of events | Events Execution Time(Seconds) |
|---|---|---|
| Chunk A | 5000 | 1.18 |
| Chunk B | 10000 | 3.64 |
| Chunk C | 15000 | 4.87 |
| Chunk D | 20000 | 11.27 |

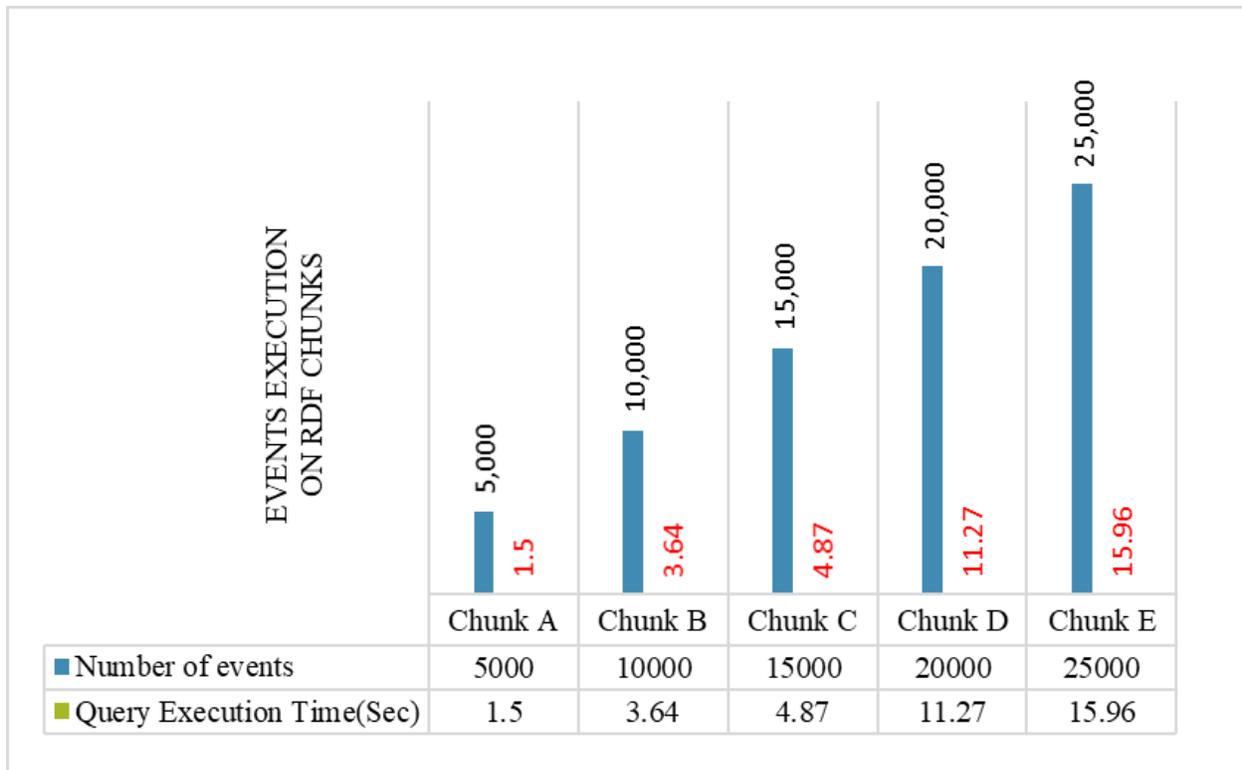

Fig.19 : Event processing time using chunks of RDF

**5 Conclusion and Future Work**

Air quality in smart cities is one of the major challenges that causes major health issues for humans. It is essential to monitor so that appropriate measures can be considered for occupants. Nobody can avoid the harm that high AQI causes for any smart city premises without taking it into consideration. Although we can diminish the consequences by monitoring air quality in real time and generating timely decision support. In this research paper, we are focused on proposing an integrated, novel approach for RCEPAQM using the CPCB dataset. The proposed model works with a real-time approach for query-based event processing to support all the stakeholders. The future work may be inclined towards developing rules based on fuzzy logic, artificial intelligence, and Web technologies stakeholders. For experimental analysis, the Jena framework streams the RDF data. A Siddhi CEP engine is used for correlating and executing these events for effective AQI monitoring using rules based on CPCB standard parameters and decision tree rules. For evaluating the proposed work, various queries are executed, and its execution time is observed by considering different scenarios to justify the accuracy and usefulness of the proposed work. Simultaneously, a large number of events are passed through different RDF chunks to validate the model. The future work may be inclined towards developing rules based on fuzzy logic. By combining artificial intelligence with big data technology and the semantic web, we can further improve the model's performance.